\documentclass[aps,nofootinbib,eqsecnum,prd,twocolumn,superscriptaddress]{revtex4-2}
\usepackage{epsfig}  
\input{epsf}
\usepackage[dvipsnames]{xcolor}
\usepackage[colorlinks=true,linktocpage=true,linkcolor=blue,citecolor=blue,urlcolor=blue]{hyperref}
\usepackage{orcidlink}
\usepackage{graphicx}
\usepackage{amsmath}
\usepackage{amssymb}
\usepackage{marginnote}
\usepackage{booktabs}
\usepackage{multirow}
\usepackage{bm}
\usepackage[utf8]{inputenc}
\usepackage{pgfplots}
\DeclareUnicodeCharacter{2212}{−}
\usepgfplotslibrary{groupplots,dateplot}
\usetikzlibrary{patterns,shapes.arrows}
\pgfplotsset{compat=newest,every axis plot/.append style={line width=2.8pt}}
\def\beq{\begin{equation}}
\def\eeq{\end{equation}}
\def\beqa{\begin{eqnarray}}
\def\eeqa{\end{eqnarray}}
\def\nn{\nonumber}
\usepackage{float}
\begin{document}

\title{Spontaneous Spin Polarization in Dense Quark Matter Reexamined}

\author{Carlisson Miller\orcidlink{0000-0003-4185-8356}} 
\email{cmc.pereira@unesp.br}
\affiliation{Instituto de F\'{\i}sica Te\'orica, Universidade Estadual Paulista, Rua Dr. Bento Teobaldo Ferraz, 271 - Bloco II, 01140-070 S\~ao Paulo, SP, Brazil}

\author{Rodolfo Casana\orcidlink{0000-0003-1461-3038}}
\email{rodolfo.casana@ufma.br}
\affiliation{Programa de P\'os-gradua\c{c}\~ao em F\'isica, Universidade Federal do Maranh\~ao,\\
 65080-805 S\~ao Lu\'is, MA, Brazil}

\author{Manoel M. Ferreira\orcidlink{0000-0002-4691-8090}}
\email{manoel.messias@ufma.br}
\affiliation{Programa de P\'os-gradua\c{c}\~ao em F\'isica, Universidade Federal do Maranh\~ao,\\
 65080-805 S\~ao Lu\'is, MA, Brazil}

\author{Gast{\~a}o Krein \orcidlink{0000-0003-1713-8578}} 
\email{gastao.krein@unesp.br}
\affiliation{Instituto de F\'{\i}sica Te\'orica, Universidade Estadual Paulista, Rua Dr. Bento Teobaldo Ferraz, 271 - Bloco II, 01140-070 S\~ao Paulo, SP, Brazil}

%
\begin{abstract}
The phase structure of strongly interacting matter at high baryon density remains one of the major open problems in QCD, with potentially important consequences for the internal structure and magnetic properties of neutron stars. Among the possible phases that may emerge under such conditions, a quark spin-polarized (QSP) phase is of particular interest because of its potential impact on both the equation of state and spontaneous magnetization. We reexamine the spontaneous emergence of a QSP phase in cold and dense quark matter within the two-flavor Nambu–Jona-Lasinio model with tensor interactions, focusing on its sensitivity to regularization. We contrast the traditional regularization scheme (TRS), in which a sharp momentum cutoff is applied to vacuum and medium contributions, with the medium separation scheme (MSS), where algebraic subtractions isolate ultraviolet-divergent vacuum terms while finite medium-dependent phase-space contributions remain cutoff free. The two prescriptions lead to qualitatively different physics. In TRS, the QSP condensate eventually collapses at high chemical potential, producing nonmonotonic thermodynamic behavior and, for some couplings, violations of causality. In MSS, by contrast, the QSP condensate increases monotonically with density, the equation of state remains causal, and the associated spontaneous magnetization grows toward high density. The persistence of spin polarization in this regime opens interesting possibilities for a microscopic contribution to the strong magnetic fields of compact stars, including magnetars. Using an exploratory hybrid-star construction, we further find that the QSP phase produces significant changes in the equation of state, stability trends, and mass-radius relations. While the stellar analysis is exploratory, the results demonstrate the potential impact of a persistent spin-polarized quark phase on neutron-star physics and highlight the importance of a consistent treatment of medium phase space.
\end{abstract}

\maketitle

%
\section{Introduction}

The properties of quantum chromodynamics (QCD) matter at high baryon density and low temperature remain poorly understood. The main reason for this lack of understanding is the so-called sign problem, which severely limits the applicability of first-principles calculations using large-scale lattice Monte Carlo simulations~\cite{Borsanyi:2010bp,Ejiri:2012rr}. Understanding strongly interacting matter under such conditions is of significant astrophysical interest, as it is directly related to the internal structure of neutron stars and possibly other compact stellar objects. Several exotic phases have been proposed to emerge in this regime, including color-superconducting and ferromagnetic phases, which may play a relevant role in these environments~\cite{Alford:2007xm,Rajagopal:2000ff,Itoh:1970uw,Ma:2022fmu}.

These possibilities have attracted considerable attention from both the theoretical and observational communities. Observations of massive neutron stars, such as PSR J0740+6620 and PSR J1614–2230, as well as gravitational wave signals from neutron star mergers, have placed stringent constraints on the equation of state (EoS) at supranuclear densities~\cite{Baym:2017whm,Blaschke:2018mqw,Lattimer:2021emm,MUSES:2023hyz,Chatziioannou:2024tjq}. Furthermore, the existence of strongly magnetized neutron stars, known as magnetars, adds an additional level of complexity to the problem~\cite{Thompson:1995gw,Chanmugam:1992rz,Harding:2006qn,Turolla:2015mwa,Esposito:2018gvp}. These stars exhibit surface magnetic field strengths on the order of $10^{14}$–$10^{15}$ G, and theoretical models indicate that the internal fields in their cores may attain even larger magnitudes. Nevertheless, the physical mechanism responsible for the generation and long-term maintenance of such extreme magnetic fields remains elusive. One proposed scenario~\cite{Tatsumi:1999ab,Tatsumi:2005ys,Tatsumi:2003bk,Nakano:2003rd}, which is the subject of the present work, is that the magnetic field has a microscopic origin, potentially arising from spin polarization of quarks in dense quark matter.

Given the aforementioned difficulties with lattice QCD, effective models of QCD have played an essential role in exploring the behavior of quark matter under extreme conditions. Among these, the Nambu–Jona-Lasinio (NJL) model~\cite{Nambu:1961tp,Nambu:1961fr} stands out due to its ability to capture essential nonperturbative features of QCD, especially chiral symmetry breaking and its restoration at high densities or temperatures~\cite{Klevansky:1992qe,Hatsuda:1994pi,Buballa:2003qv}. The NJL model, though it lacks confinement, serves as a valuable theoretical laboratory for studying dynamical mass generation and the formation of various condensates. In particular, the spontaneous emergence of a spin-polarized phase of quark matter, commonly referred to as ferromagnetic quark matter, has been extensively explored in the literature within this framework; for representative studies, see Refs.~\cite{Maedan:2006ib,Tsue:2012nx,Tsue:2014tra,Matsuoka:2016spx,Matsuoka:2018huq,Morimoto:2020fkg,Niu:2025zgm}. 

A recurrent finding in studies based on the NJL model is that, at sufficiently high quark densities, the spin-polarized phase and the associated spontaneously generated magnetic moment disappear, a result that is challenging to interpret physically. In the present work, we reexamine this question in light of the regulation of ultraviolet divergences in the model. As is well known, a critical component of any calculation within the NJL framework is the choice of an appropriate regularization scheme~\cite{Buballa:2003qv,Avancini:2020xqe,Buballa:2020nsi} owing to the intrinsic nonrenormalizability of the model. Traditionally, a sharp cutoff or similar regulator is used with divergent momentum integrals in the model, even when the integrands of those integrals involve medium-dependent contributions. Although this procedure is technically straightforward and is extensively employed in the literature, it might generate nonphysical effects in terms that intertwine vacuum and medium contributions within ultraviolet-divergent integrals. Vacuum–medium mixing may induce spurious dependencies of physical observables on temperature, chemical potential, or external background fields~\cite{Farias:2005cr,Farias:2016let,Coppola:2017edn,Allen:2015paa,Kohyama:2015hix,Kohyama:2016fif}. For example, the resulting equation of state may violate causality in certain density regimes, which is clearly physically unacceptable.  

As argued in the following, the situation with the spin-polarized phase is similar to the color-superconducting gap in quark matter: the gap vanishes for chemical potentials comparable to the ultraviolet cutoff scale~\cite{Farias:2005cr}, in contradiction with QCD expectations~\cite{Son:1998uk}, according to which the gap remains finite and increases with the chemical potential. Such spurious effects can, however, be eliminated by using a procedure introduced in Ref.~\cite{Farias:2005cr}, later termed the medium separation scheme (MSS) in Ref.~\cite{Farias:2016let}, which enables a systematic separation of medium-dependent contributions from within divergent integrals. By untangling vacuum and medium effects, the MSS leads to physically consistent and acceptable predictions. 

Although one might argue that in an effective low-energy theory like the NJL model, all momentum integrals should strictly be truncated at the cutoff scale~$\Lambda$, even when mathematically finite, this view implicitly blurs the distinction in the role of $\Lambda$ as an ultraviolet regulator of vacuum quantum fluctuations and that of a physical boundary on medium phase space. The view here is that the cutoff $\Lambda$ is introduced to regularize divergent vacuum loops arising from point-like contact interactions; it parameterizes the scale of vacuum quantum fluctuations (such as chiral symmetry breaking) rather than serving as a wall restricting density-dependent states. When evaluating finite, medium-isolated contributions, such as those from quantum statistics via Fermi-Dirac distributions, they provide their own natural, physical ultraviolet damping. Enforcing a sharp cutoff $p \le \Lambda$ on medium-dependent integrals imposes an unphysical step-function boundary in momentum space, which truncates the phase space available to active states near the Fermi surface as the chemical potential approaches~$\Lambda$. This truncation is precisely what forces order parameters (such as the spin condensate $F$ or color-superconducting gaps) to collapse at high densities while simultaneously introducing boundary-driven anomalies that violate causality ($c_s^2 > 1$) and thermodynamic consistency relations. Integrating the finite, medium-isolated parts over the full momentum space prevents these regulator-induced boundary effects without adding unphysical high-energy vacuum content, provided the chemical potential remains within the domain of validity of the effective theory ($\mu_q \le \Lambda$).

In this paper, we consider the SU(2) version of the NJL model augmented by a tensor interaction that can trigger a spin polarized phase. We perform a comparative analysis using both the traditional regularization scheme (TRS) and MSS frameworks, and we demonstrate the implications for the existence and properties of a spin-polarized phase in high density quark matter. We also examine how the MSS-regularized spin-polarized phase affects the equation of state and the mass-radius relation of compact stars. 

The present two-flavor treatment should be regarded as a controlled exploratory study of the regularization issue. At sufficiently high densities, strange quarks and additional pairing channels are expected to become relevant and may modify the equation of state, the spin-polarized phase, and the stellar mass-radius relation \cite{Rajagopal:2000wf,Rajagopal:2000ff,Fukushima:2008wg,Bonanno:2011ch,Morimoto:2020fkg,Kohyama:2016fif}. A realistic compact-star description should therefore include strangeness and other competing phases, which we leave for future work. We further emphasize that our analysis focuses on magnetization arising spontaneously, rather than being induced by an external spin potential that constrains the expectation value of a spin tensor; see, e.g., the review in Ref.~\cite{Becattini:2022zvf} and references therein for a detailed discussion. References~\cite{Farias:2025vss} and \cite{Singha:2025bda} implement such a spin potential within the NJL model and the linear sigma model with quarks, respectively.

The structure of the paper is as follows. In Sec.~\ref{sec:MODEL}, we present the SU(2) NJL model with a tensor interaction and the relevant thermodynamic formalism. In Sec.~\ref{sec:REGULARIZATION}, we introduce the regularization schemes and outline the implementation of the MSS. In Sec.~\ref{sec:RESULTS}, we analyze and compare the physical consequences of each scheme for the spin-polarized phase, the equation of state, and the mass-radius relation of compact stars. Finally, in Sec.~\ref{sec:conclusion}, we summarize our findings and discuss possible extensions of this work. 

%
\section{Thermodynamics of the model}
\label{sec:MODEL}

In this study, we employ the following $SU(2)$-fla\-vor Lagrangian density:
\begin{eqnarray}
\mathcal{L} &=&  \bar\psi \left( i\gamma^{\mu}\partial_{\mu} - m \right) \psi
+ G_S \left[ \left(\bar\psi \psi \right)^2 + \left(\bar\psi i \gamma^{5} \vec{\tau} \psi \right)^2 \right] \nonumber \\
&-& \frac{G_T}{4} \Big[
\left(\bar\psi \gamma^{\mu} \gamma^{\nu} \vec{\tau} \psi \right) \cdot
\left(\bar\psi \gamma_{\mu} \gamma_{\nu} \vec{\tau} \psi \right) \nonumber \\
&+& \left(\bar\psi i \gamma^{5} \gamma^{\mu} \gamma^{\nu} \psi \right)
\left(\bar\psi i \gamma^{5} \gamma_{\mu} \gamma_{\nu} \psi \right)  \Big]  ,
\label{MODEL_EQ2}
\end{eqnarray}
where $\psi$ denotes the four-component Dirac spinor field describing quarks and carrying both flavor and color degrees of freedom. Since the model is nonrenormalizable, the choice of regularization scheme constitutes an integral part of its definition; in the present work, we adopt a three-dimensional momentum cutoff~$\Lambda$. We also work in the chiral limit ($m=0$). Therefore, the free parameters of the model are the effective coupling constants $G_S$ and $G_T$, and the cutoff~$\Lambda$. Following earlier investigations, we restrict ourselves to spin polarization along the $z$-axis, so that the corresponding spin operator is given by:
\begin{eqnarray}
\Sigma_{3} = -i \gamma^{1}\gamma^{2} =
\begin{pmatrix}
\sigma_3 & 0 \\
0 & \sigma_3
\end{pmatrix}, \qquad \sigma_3 = \begin{pmatrix}
1 & 0 \\
0 & -1
\end{pmatrix} .
\label{MODEL_EQ4}
\end{eqnarray}

We will work in the mean field approximation. In this approximation, the grand thermodynamic potential can be written as $\Omega = \Omega_0 + \Omega_q + \Omega_e$, where~\cite{Matsuoka:2016spx, Matsuoka:2016ufr}: 
\begin{widetext}
\begin{eqnarray}
\Omega_0 &=& \frac{\sigma^2}{4G_{S}} + \frac{F^2}{2G_{T}}, \label{Omega0} \\
\Omega_q &=& 
- N_c T 
\sum_{\eta=\pm 1} \sum_{f=u,d} \int \frac{d^3 p}{(2\pi)^3}\Biggl[ 
\ln\left( 1 + e^{-\frac{E^{(\eta)}_p + \mu_f}{T}} \right)  
+ \ln\left( 1 + e^{-\frac{E_{p}^{(\eta)} - \mu_f }{T}} \right) \Biggr] 
- N_c N_f \sum_{n=\pm 1}
\int \frac{d^3 p}{(2\pi)^3} 
E_p^{(\eta)}, \label{Omegaq} \\
\Omega_e &=& - \frac{\mu_e^4}{12\pi^2},
\label{Omegae}
\end{eqnarray}
where $\sigma = -2 G_{S} \langle \bar\psi \psi
\rangle$, $F = F_3 = - G_{T} \langle \bar\psi \Sigma_{3} \tau_3 \psi \rangle$ is the spin condensate, $N_c=3$ and $N_f = 2$ are the numbers of colors and flavors, and $E_{p}^{(\eta)}$ is the single-particle energy: 
\begin{eqnarray}
E_{p}^{(\eta)} = \sqrt{p^{2}_{3} + \left( E_{p_t} + \eta F \right)^2},
\label{MODEL_EQ8}
\end{eqnarray}
with $E_{p_t}=\sqrt{p^{2}_t + M^2}$ where $p_t = \sqrt{p^{2}_{1}+p^{2}_{2}}$,  $M=\sigma$ is the dynamical quark mass, and $\eta=\pm 1$ are the eigenvalues of the $\tau_3$. It should be emphasized that, although we work in the SU(2) flavor-symmetric limit, flavor-dependent chemical potentials must still be introduced. This is necessary in order to impose $\beta$-equilibrium conditions, which are required when studying the mass–radius relation of a star. Due to the breaking of rotational symmetry, it is convenient to express the three-dimensional integration element as follows:
\begin{eqnarray}
\int \frac{d^3 p}{(2\pi)^3} \, \Bigl( \cdots \Bigr)
=  \frac{1}{2\pi^2}
\int^{\Lambda}_{0} dp_{t}
p_{t}
\int^{\sqrt{\Lambda^2 - p_{t}^{2}}}_{0} dp_{3} \, \Bigl( \cdots \Bigr).
\end{eqnarray}
The values of the condensates $\sigma$ and $F$ are determined by gap equations that follow from minimizing the thermodynamic potential:
\begin{eqnarray}
\frac{\sigma}{2G_S} &=& N_c  \sum_{\eta=\pm 1} \sum_{f=u,d} \int \frac{d^3 p}{(2\pi)^3} \, \frac{M}{E^{(\eta)}_p}
\frac{E_{p_t} + \eta F}{E_{p_t}} 
\left( 1 - n^f_{+}(E^{(\eta)}_p) - n^f_{-}(E^{(\eta)}_p) \right) ,
\label{chiral_gap} \\ 
\frac{F}{G_T} &=& N_c \sum_{\eta=\pm 1} \sum_{f=u,d} \int \frac{d^3 p}{(2\pi)^3} \, \eta \frac{E_{p_t} + \eta F}{E^{(\eta)}_{p}} 
\left( 1 - n^f_{+}(E^{(\eta)}_p) - n^f_{-}(E^{(\eta)}_p) 
\right),
\label{qsp_gap}
\end{eqnarray}
\end{widetext}
where $E_{p_t} = \sqrt{p_{t}^2 + M^2}$ and here, $n_{\pm}$ denote the quark and anti-quark Fermi distribution
functions: 
\begin{eqnarray}
n^{f}_{\pm}(E^{(\eta)}_p) = 
\frac{1}{e^{(E^{(\eta)}_p \pm \mu_{f})/T} + 1 }.
\end{eqnarray}

In this study, we focus on the thermodynamic properties of pressure $P = -\Omega$, the quark charge density $\rho$, the entropy density $s$ and energy density $\epsilon$. These properties are crucial as they define the system's equation of state (EOS), with pressure being a key factor determining the system's behavior. Thus, we have 
\begin{eqnarray} 
P_N = P(T, \mu) - P(0,0) , 
\label{press_renor}
\end{eqnarray}
\begin{eqnarray} 
\epsilon = -P_N + T s + \sum_{i=u,d,e} \mu_i \rho_i \left(\mu_i\right) ,
\label{energy}
\end{eqnarray}
where we have subtracted the zero point pressure in order to obtain a normalized pressure, $P_N$. To avoid excessive notation, we will omit the normalization index in $P_N$ from now on.  is calculated by determining its final term in Eq.~\eqref{Omegaq}. This is achieved by integrating the right-hand side with respect to the field $F$ in Eq.~\eqref{REGULARIZATION_EQ3}. The explicit form of the entropy 
density, obtained by taking the derivative of the thermodynamic potential with respect to the temperature, is 
\begin{eqnarray}
s &=& N_c \sum_{\eta,f, \alpha} \int \frac{d^3 p}{(2\pi)^3} 
\Biggl[ 
\ln\left( 1 + e^{-(E^{(\eta)}_p +\alpha \mu_{f})/T} \right)  \nonumber \\
&+&  \frac{(E^{(\eta)}_p +\alpha \mu_{f})}{T} n^{f}_{\alpha}(E^{(\eta)}_p) \Biggr],
\end{eqnarray}
where the index $\alpha =\pm 1$ labels particle and antiparticle excitations, with $\alpha=-1$ for quarks 
and $\alpha=+1$ for antiquarks. Finally, the quark density~$\rho$ and the square of the speed of sound~$c^2_s$ are given by
\begin{align}
\rho &= N_c \sum_{\eta, f} \int \frac{d^3 p}{(2\pi)^3}
\left[n^{f}_{-}(E^{(\eta)}_p) - n^{f}_{+}(E^{(\eta)}_p) \right], 
\label{density} \\ 
c^2_s &= \frac{\partial P}{\partial\epsilon} .
\label{sound_velocity}
\end{align}

To investigate the influence of the MSS-regularized spin-polarized phase on the mass–radius relation of neutron stars, it is necessary to impose $\beta$-equilibrium in the stellar core as dictated by weak-interaction processes:
\begin{eqnarray}
d \rightarrow u + e + \bar{\nu}_e , \quad 
u + e \rightarrow d + \nu_e .
\label{beta_reaction}
\end{eqnarray}
Therefore, the following $\beta$-equilibrium condition should be satisfied
\begin{eqnarray}
\mu_d = \mu_u + \mu_e .
\label{beta_equilibrium}
\end{eqnarray}
where $\mu_{u(d)}$ are the chemical potentials for quarks up and down, respectively, and $\mu_e$ is the chemical potential for the electrons. For electrically neutral quark matter, the charge quark densities $\rho_{u(d)}$, for up and down quarks respectively, are related to the electron density $\rho_{e}$ via:
\begin{eqnarray}
\frac{2}{3}\rho_u 
- \frac{1}{3}\rho_d
- \rho_e
= 0 ,
\label{Charge_Neutrality}
\end{eqnarray}
where each charge density depends on its respective chemical potential, and the baryonic charge density is defined by 
\begin{eqnarray}
\rho_B
= \frac{1}{3}\left(
\rho_u + \rho_d \right).
\label{Baryon_Neutrality}
\end{eqnarray}
Due to the relativistic speeds of the electrons, their particle number density can be described 
as a gas of noninteracting electrons $\rho_e = \mu_e^{3}/(3\pi^2)$. Given the constraints of $\beta$-equilibrium and charge-neutrality, only one independent chemical potential remains. We select $\mu_u$, while the other chemical potentials $\mu_d$ and $\mu_e$ can be treated as functions of $\mu_u$.

%
\section{REGULARIZATION SCHEMES}
\label{sec:REGULARIZATION}

The quark kinetic contribution to the grand thermodynamic potential, last term in Eq.~\eqref{Omegaq}, is ultraviolet divergent and requires regularization.  This term depends indirectly on $\mu$ and $T$ through its dependence on $M$ and $F$, thereby entangling vacuum contributions with medium-dependent components. As mentioned previously, using the TRS leads to vanishing spin polarization and spontaneous magnetization in high density quark matter. We examine this issue using the MSS procedure to separate the dependencies of vacuum-related terms from the medium effects in the current problem. This discussion starts with the broader scenario, the physical case with a nonzero current quark mass~$m$, and subsequently extends to the chiral limit for the MSS regularized integrals.

Firstly, let us focus on the UV divergent term in the gap equation for $F$, the first term in Eq.~\eqref{qsp_gap}. This contribution can be rewritten as a Euclidean integral:
\begin{eqnarray}
\int \frac{d^3 p}{(2\pi)^3} 
\frac{H(p_{t})}{E^{(\eta)}_{p}} = 2 \int\frac{d^4 p}{(2\pi)^4} 
\frac{ H(p_{t}) }{p_{4}^2 + E^{(\eta) 2}_{p}} ,
\label{REGULARIZATION_EQ1}
\end{eqnarray}
where we have defined $H(p_{t}) = E(p_{t}) + \eta F$. To explicitly subtract the vacuum contribution from this integral, we use the following identity: 
\begin{align}
\frac{1}{ (p_{4} )^{2} + E_{p}^{(\eta)2} }
&= \frac{1}{p_{4}^{2} + \bm{p}^2 + M^2_0} \nn \\
&+ \frac{\Sigma - 2\eta
E_{p_t} F}{(p^2_4 + \bm{p}^2 + M^2_0)
(p_{4}^{2} + E_{p}^{(\eta)2})},
\label{REGULARIZATION_EQ2}
\end{align}
where $M_0$ is the dynamical quark mass in vacuum and $\Sigma = M_0^2 - F^2$. Next, repeating the subtraction four times using this same identity and writing $p^2_4 + \bm{p}^2 + M^2_0 = p^2 + M^2_0$, we obtain: 
\begin{widetext} 
\begin{eqnarray}
\hspace{-1.0cm}\frac{1}{ (p_{4} )^{2} + E_{p}^{(\eta)2} }
&=& \frac{1}{p^{2} + M^2_0}
+ \frac{\Sigma - 2\eta E_{p_t} F}{\left(p^{2} + M^2_0\right)^2 } 
+ \frac{(\Sigma - 2\eta E_{p_t} F)^2}{\left(p^{2} + M^2_0 \right)^3}  
+ \frac{(\Sigma - 2\eta E_{p_t} F)^3}{\left(p^{2} + M^2_0 \right)^4} 
+ \frac{(\Sigma - 2\eta E_{p_t} F)^4}{\left(p^{2} + M^2_0 \right)^4
\left(p_{4}^{2} + E_{p}^{(\eta)2}\right)}. 
\label{REGULARIZATION_EQ3}
\end{eqnarray}
\end{widetext} 
Upon substituting this expression into Eq.~\eqref{REGULARIZATION_EQ1}, the first term corresponds to the gap equation that determines the dynamically generated quark mass~$M_0$ in the vacuum, the last term gives a finite integral, whereas the second, third, and fourth terms lead to divergent integrals.  The divergent integrals mirror those encountered in the vacuum analysis, in the sense that their integrands depend solely on the constituent vacuum quark mass~$M_0$ and are independent of $T$ and $\mu$. This follows from the fact that $F = F(T,\mu)$ can be factored out of these integrals.

\begin{figure*}
\begin{center}
\includegraphics[scale=0.5]{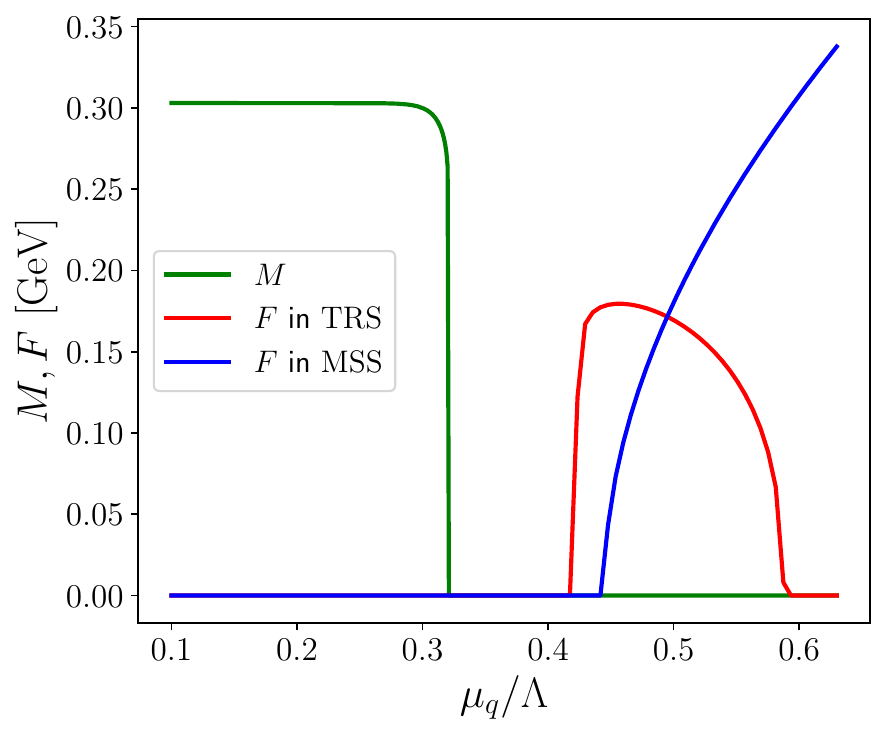}\hspace{0.5cm}
\includegraphics[scale=0.5]{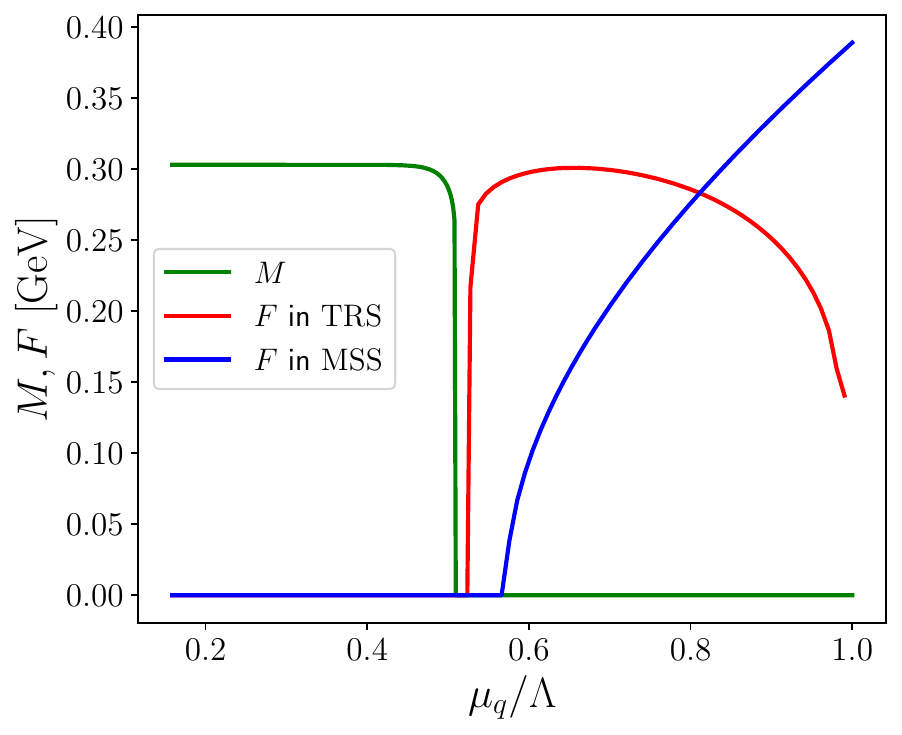}
\end{center}
\vspace{-0.5cm}
\caption{Dynamically generated quark mass~$M$ and spin polarization $F$ as a function of quark chemical potential (normalized by $\Lambda$) in the chiral limit. Left panel: $G_T = 11~\text{GeV}^{-2}$. Right panel: 
$G_T = 14~\text{GeV}^{-2}$. }
\label{fig:M-F}
\end{figure*}

At this stage, the MSS methodology departs from the conventional TRS in that Eq.~\eqref{REGULARIZATION_EQ1} is evaluated directly with a momentum cutoff. Within the MSS framework, Eq.~\eqref{REGULARIZATION_EQ3} is first substituted into Eq.~\eqref{REGULARIZATION_EQ1}, and only the divergent integrals that result from this manipulation are integrated using a cutoff. An analogous treatment is applied to the integrals entering the thermodynamic quantities. This difference in the regularization procedure can substantially alter the resulting solutions, particularly those related to spin polarization.

The spin polarization phase, described by $F$, is established self-consistently through the gap equation 
in Eq.~\eqref{qsp_gap}. Using the identity in Eq.~\eqref{REGULARIZATION_EQ3}, this gap equation can be written as:
\begin{equation} 
\frac{F}{N_c N_f G_T} =  \sum_{\eta} \Bigl[T_{\rm fin}(\eta,F,M,M_0)  + T_{\rm div}(\eta,F,M,M_0)\Bigr], 
\label{F-final}
\end{equation}
where 
\begin{widetext}
\begin{align}
T_{\rm fin} (\eta, F, M_0) &=  2 \eta \int\frac{d^4 p}{(2\pi)^4} 
\frac{H(p_t)}{\left(p^2 + M^2_0\right)^4} \left[ 
\frac{\left(\Sigma - 2 \eta E_{p_t} F\right)^4}{p_{4}^{2} + E_{p}^{(\eta)2}} 
+ 12 E_{p_t}^2 F^2\Sigma 
- 6\eta \Sigma^2 E_{p_t} F  + \Sigma^3  \right] \nonumber \\
&+ \eta \int \frac{d^3 p}{(2\pi)^3} \, \frac{H(p_t)}{E^{(\eta)}_{p}} 
\Big[ 
 n_{+}(E^{(\eta)}_p) + n_{-}(E^{(\eta)}_p) 
\Big] , \\
T_{\rm div}(\eta, F, M_0) &= 2 \eta \int\frac{d^4 p}{(2\pi)^4} \, \frac{H(p_t)}{p^2 + M^2_0}\left[ 1 + 
\frac{\Sigma - 2\eta E_{p_t} F}{p^{2} + M^2_0}
+ \frac{\left(\Sigma - 2\eta E_{p_t} F\right)^2}{(p^{2} + M^2_0)^2} 
-  8\eta F^3  \frac{E_{p_t}^3}{(p^2 + M^2_0)^3}  \right].
\end{align} 
\end{widetext}
As previously noted, the spin-polarization field~$F$ appears exclusively in the numerator in all divergent integrals. Since it does not depend on the integration variables, it~can be factored out of the integrals, which then depend only on the vacuum constituent quarks mass~$M_0$.

%
\section{NUMERICAL RESULTS}
\label{sec:RESULTS}

In this section, we present the numerical results and perform a comparative analysis of the outcomes obtained using the MSS (medium separation scheme) and the TRS (traditional regularization scheme). We employ the same parameter values for $G_S$ and $\Lambda$ as those adopted in Ref.~\cite{Matsuoka:2018huq}, specifically $G_S = 5.5~\text{GeV}^{-2}$ and $\Lambda = 0.63~\text{GeV}$. As discussed in detail in Ref.~\cite{Matsuoka:2016ufr}, the value of the tensor coupling $G_T$ remains insufficiently constrained; therefore, we treat $G_T$ as a free parameter.

\subsection{Dynamical quark mass and spin polarized phase}

We determined the constituent quark mass~$M$ and the spin polarization~$F$ by solving the corresponding gap equations under imposed chemical equilibrium constraints. In this study, we confine our analysis to values of the chemical potential that are compatible with interpreting the model as an effective, nonrenormalizable field theory characterized by an intrinsic ultraviolet cutoff scale~$\Lambda$, beyond which its predictive capacity is intrinsically constrained. Although there are ways to extend the applicability of the model beyond $\Lambda$ by making the couplings and cutoff~$T$ and $\mu_q$ dependent~\cite{Casalbuoni:2003cs,Farias:2006cs}, to avoid misinterpretations regarding the MSS, we accordingly work within the condition $\mu_q \lesssim \Lambda$.

Figure~\ref{fig:M-F} displays the $\mu$-dependence of~$M$~and~$F$ at $T = 0.01~\text{GeV}$, for coupling strengths $G_T=11~\text{GeV}^{-2}$ (left panel) and $G_T=14~\text{GeV}^{-2}$ (right panel). The mass~$M$ is the same in TRS and MSS. A~first notable observation is that, as functions of $\mu_q$, the chirally broken phase and the ferromagnetic phase do not overlap; however, for the larger coupling, they approach each other closely. Alternative parameter choices can produce an overlap between these phases, but since no additional qualitative features emerge, we restrict the subsequent analysis to the parameter sets used above. Furthermore, the QSP predicted within the MSS displays qualitatively different behavior from the corresponding TRS result: in the MSS, the QSP develops continuously and increases monotonically with increasing chemical potential, whereas in the TRS it emerges discontinuously and then decreases rapidly, vanishing at larger $\mu$.

In the following, we quantitatively evaluate how discrepancies between the order parameters predicted by TRS and MSS alter thermodynamic observables, and we examine the possible implications for neutron-star structure. We further show that MSS enables the dynamical generation of a sizable magnetic moment, thereby potentially providing a mechanism for the strong magnetic fields inferred in magnetars. 

%
\subsection{Spin-polarized phase and thermodynamic properties of quark matter}

\begin{figure}[htb]
\begin{center}
\includegraphics[scale=0.5]{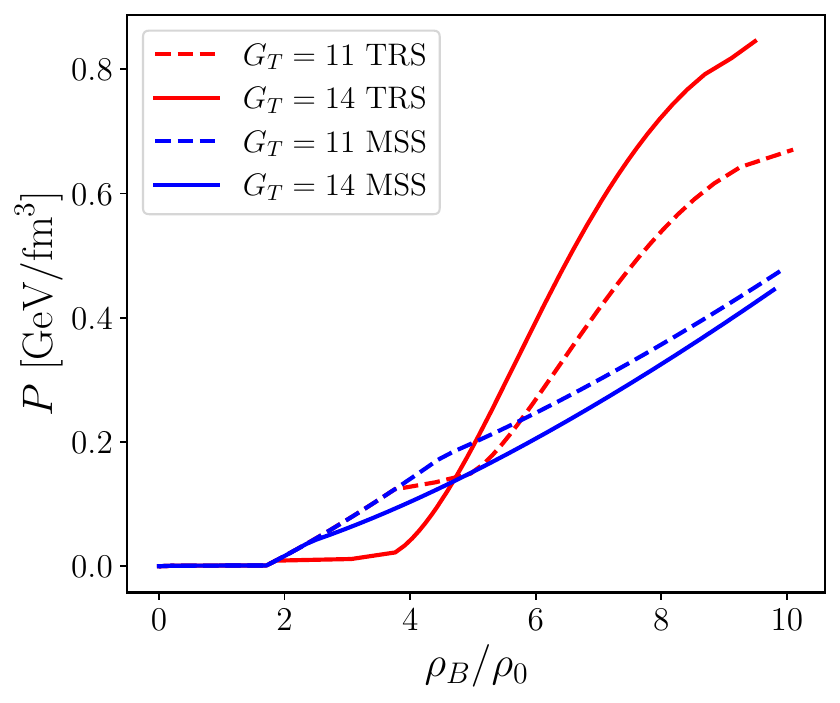}\hspace{2.0cm}
\includegraphics[scale=0.5]{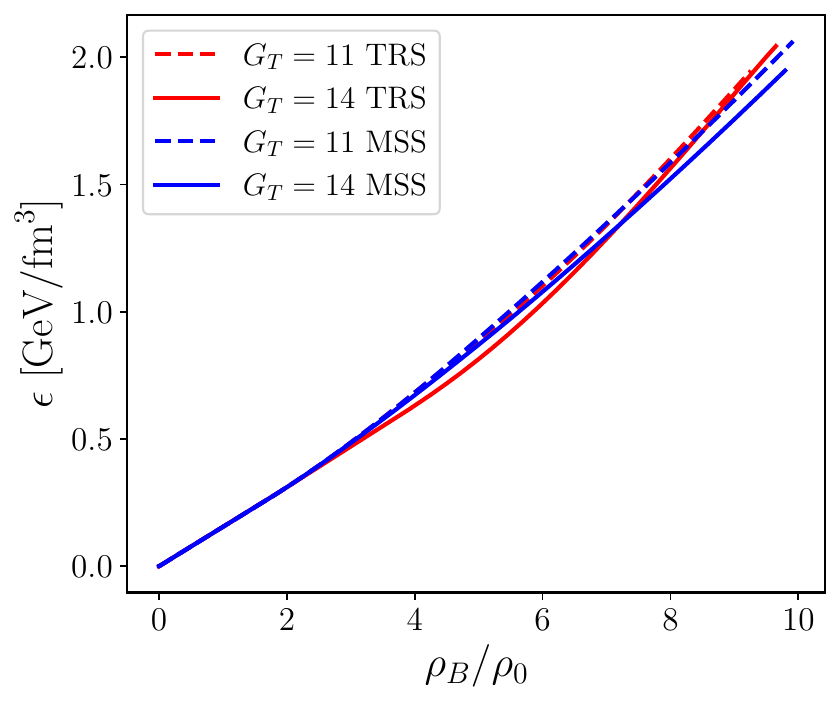}
\end{center}
\vspace{-0.5cm}
\caption{Pressure as a function of the baryon density normalized by $\rho_0 = 0.17~\rm{fm}^{-1}$ in TRS and MSS.}
\label{fig:Peps_rho}
\end{figure}

Figure~\ref{fig:Peps_rho} presents the pressure (top panel) and energy density (lower panel) as functions of baryon density. The results indicate that when changing from the TRS to the MSS, the pressure exhibits a more pronounced sensitivity than the energy density. For $\rho_B \gtrsim 4\rho_0$, the departure from the TRS prediction becomes significant, and the effect is enhanced for $G_T=14~\text{GeV}^{-2}$. This behavior has direct implications for the quark-phase equation of state (EOS).

\begin{figure*}[htb]
\begin{center}
\includegraphics[scale=0.5]{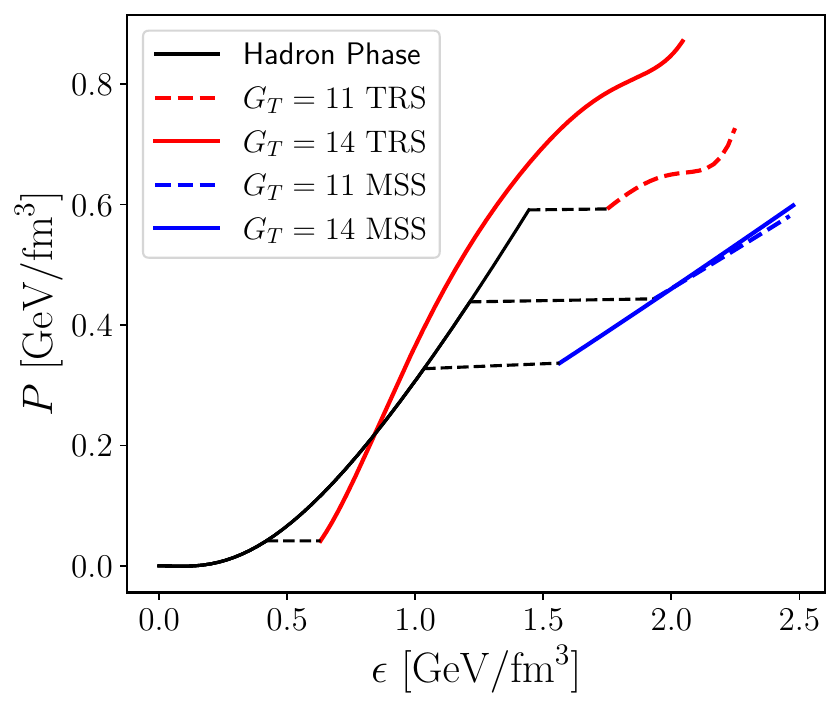}\hspace{0.5cm}
\includegraphics[scale=0.5]{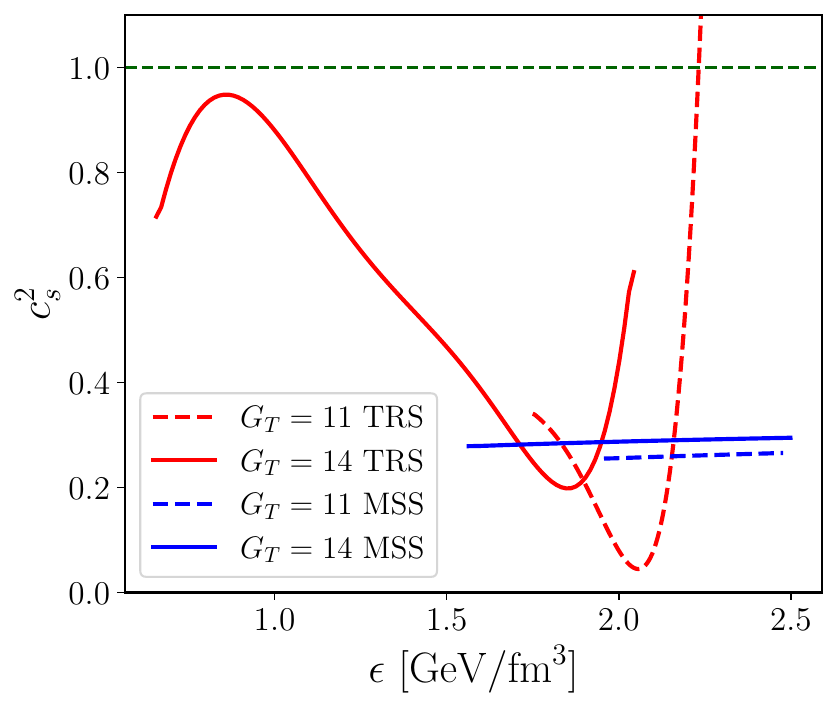}
\end{center}
\vspace{-0.5cm}
\caption{Equation of state in TRS and MSS.}
\label{P_ener_vs}
\end{figure*}

Figure~\ref{P_ener_vs} displays a hybrid EOS featuring a mixed hadron–quark phase, which is discussed in the following paragraph; the quark phase manifests at higher baryon densities. Specifically, for $\epsilon \gtrsim 1.2~\text{GeV}/\text{fm}^{3}$, the QSP phase obtained with the MSS yields a lower pressure than that predicted by the TRS for both considered values of the coupling $G_T$, implying a softer EOS.

A softer equation of state (EOS) typically reduces the maximum gravitational mass of neutron stars, making it challenging to satisfy the observational constraint of $\sim 2\,M_\odot$~\cite{Baym:2017whm}. To assess the potential implications of a quark spin-polarized (QSP) phase for neutron-star structure, we investigate a hybrid EOS incorporating a hadron–quark phase transition. Such stars would have an inner core made up of quark matter and an outer core composed of hadrons. For the hadronic sector, we adopt the non-linear Walecka model, in which nucleons interact via the exchange of scalar $\sigma$ and vector $\omega$ mesons~\cite{Schmitt:2010pn}. The non-linear extension is used mainly because it addresses a major limitation of the linear formulation, which yields an unrealistically stiff nuclear matter EOS. We emphasize that our aim in using this framework is not to provide a state-of-the-art quantitative description of neutron stars, but rather to obtain qualitative insight into the microscopic effects introduced by the spin-polarized phase.

The first-order phase transition between nuclear matter and quark matter is modeled via the Maxwell construction. At the transition point, phase equilibrium is imposed by enforcing equality of the baryon chemical potential and the pressure in the two phases. For energy densities beyond the coexistence point, the thermodynamically favored phase is identified as the one with the larger pressure. Figure~\ref{P_ener_vs} displays the resulting complete equation of state, presenting the pressure as a function of energy density for the methods introduced above. Within the TRS scheme, increasing the tensor coupling $G_T$ produces a pronounced modification of the transition point; in particular, for $G_{T}=14~\text{GeV}^{-2}$ the spin-polarized phase appears already at lower energy densities. In the MSS scheme, the transition point also exhibits a dependence on $G_T$, albeit with a comparatively modest shift. Moreover, the width of the transition region in energy density is sensitive to the chosen regularization procedure, with the MSS scheme yielding a wider coexistence interval than the TRS scheme.

The corresponding sound-speed results for quark matter are shown in the right panel of Fig.~\ref{P_ener_vs}. In particular, within the TRS approach, a causality violation occurs for $G_T=11~\text{GeV}^{-2}$ at $\epsilon \approx 2.2~\text{GeV}/\text{fm}^{3}$, corresponding to baryon densities of the order of $\rho_B \sim 9\rho_0$. By contrast, no such violation is found for $G_T=14~\text{GeV}^{-2}$, attributable to the stronger quark spin-polarization condensate generated at this coupling. For the MSS, causality is preserved for both coupling values. This conclusion is consistent with Ref.~\cite{Pasqualotto:2023hho}, which employed the MSS without considering a QSP phase.

%
\subsection{Spin-polarized phase and the structure of hybrid stars}

Here, we use our findings to explore the potential implications of the MSS for the properties of hybrid neutron stars. Even though the presence of a magnetic field leads to anisotropic pressure, the resulting deformation is likely to be minimal within the range of magnetic fields considered here. Therefore, we can safely assume spherical symmetry for the neutron star's structure. To~determine the properties of a neutron star, the Tolman-Oppenheimer-Volkoff (TOV) equation must be solved for a static, non-rotating, spherically symmetric star
\begin{eqnarray}
\frac{dP(r)}{dr} = 
\frac{G}{c^2} 
\left(\varepsilon(r) + P(r)\right)
\frac{M(r) + 4\pi r^3 P(r)}{r\left(r-2 G M(r)/c^2\right)},
\end{eqnarray}
where $G$ denotes the gravitational constant and
\begin{eqnarray}
\frac{dM(r)}{dr} = 
4\pi r^3 
\frac{\varepsilon(r)}{c^2},
\end{eqnarray}
where $G$ represents the gravitational constant,  $m$ is the accumulated mass (the mass within a sphere of radius~$r$), and $P$ and $\varepsilon$ are the pressure and energy density at a distance~$r$ from the star's center. This system of equations is solvable when the function $\varepsilon(P)$, i.e., the equation of state of the stellar matter, is known. We reiterate that employing the conventional Tolman–Oppenheimer–Volkoff (TOV) equation in this context constitutes an extrapolation because the emergence of the quark spin-polarized phase leads to the spontaneous breaking of rotational symmetry. Accordingly, the TOV framework is used here primarily for qualitative guidance. Numerical integration starts from the center with a specified central pressure $P(0) = P_c$, where the boundary condition $M(0) = 0$ is met. The star's radius $R$ is established by the condition $P(R) = 0$, and the star's gravitational mass is $M = M(r = R)$.

\begin{figure}[H]
\begin{center}
\includegraphics[scale=0.5]{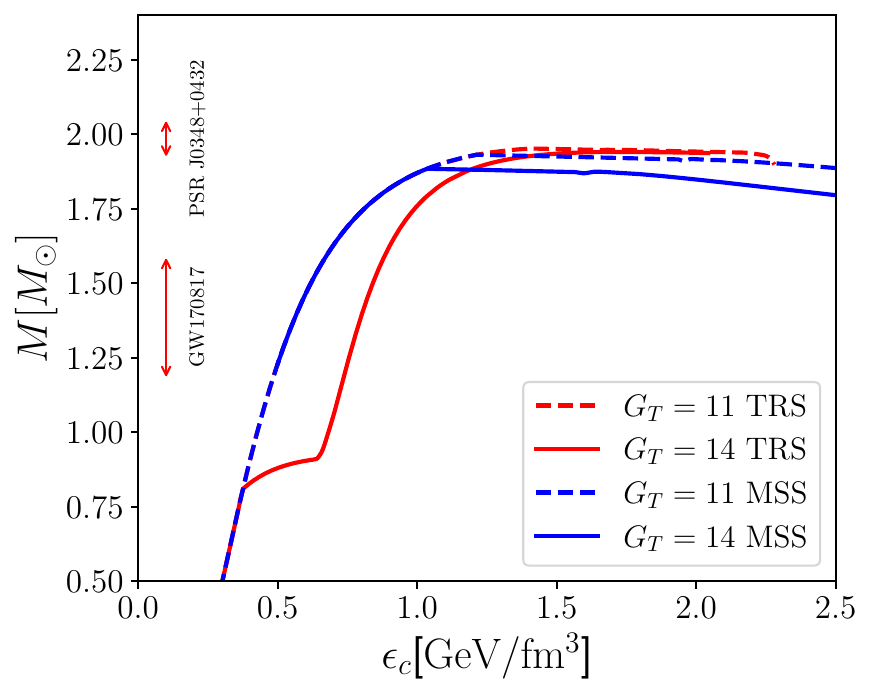}
\includegraphics[scale=0.5]{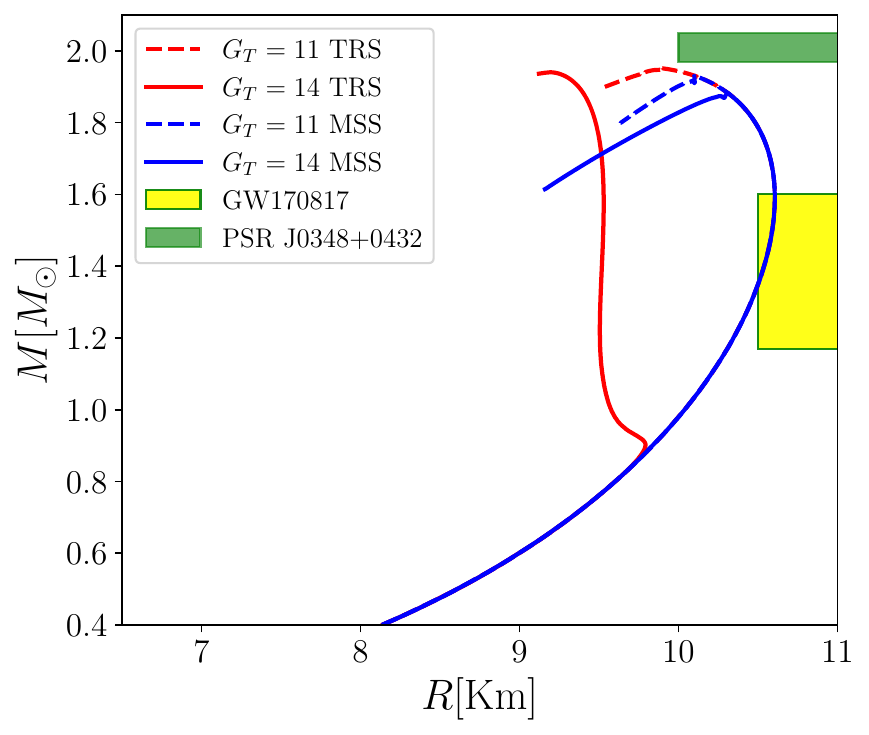}
\end{center}
\vspace{-0.5cm}
\caption{Star masses as functions of the central energy
density.}
\label{Mass_star}
\end{figure}

Figure~\ref{Mass_star} displays (in the top panel) the dependence of the gravitational mass of a compact star on its central energy density for both TRS and MSS methods. For guidance, the upper vertical arrow represents the measured mass of the pulsar PSR J0348+0432, which is $M = (2.01 \pm 0.04) M_{\odot}$\cite{Antoniadis:2013pzd}. This measurement provides a lower bound for estimating the maximum mass of neutron stars. The lower arrow represents the recent observed neutron star mass values from GW170817 are $M = (1.17 \pm 1.60) M_{\odot}$, based on gravitational wave measurements~\cite{LIGOScientific:2018cki, LIGOScientific:2017vwq}. From this plot, we can study the stability of the configurations. The condition for the stability of stars along the sequence of equilibrium configurations is based on the inequality $dM/d\varepsilon_c > 0$. For an unstable configuration, a small perturbation would be enough to cause the pressure's outward force to overcome the inward pull of gravity, resulting in the star exploding. It can be seen that the configuration in the MSS method is more sensitive to gravitational collapse than that obtained in the TRS method for both couplings, particularly at high values of central energy density. When comparing only the two configurations within the MSS method, we observe that increasing the coupling makes it more sensitive to gravitational collapse at high central energy densities.

The star's mass-radius relation changes significantly between the two methods for $G_T = 14~\text{GeV}^{-2}$, as shown in the lower panel of Fig.~\ref{Mass_star}. For the same coupling, in the TRS method, we observe that the curve bends at \(M = 1.1 M_\odot\), corresponding to the onset of the quark spin polarization phase within the star's core, and it reaches the maximum mass at \(M = 1.87 M_\odot\). In contrast, within the MSS method, since the formation of this phase occurs more smoothly, we notice only a slight bend in the mass-radius curve at \(M = 1.82 M_\odot\), which is the maximum mass achieved in this method. For the coupling $G_T = 11~\text{GeV}^{-2}$, we observe that the mass-radius curves are quite similar, but the one obtained using the MSS method bends at a lower mass compared to the TRS method.

%
\subsection{Estimation of magnetic moment}

As mentioned earlier, a non-vanishing spin polarization phase indicates the alignment of quark spins, resulting in a magnetic field due to the quark magnetic moment. Consequently, there is a possibility of spontaneous magnetization occurring in quark matter at high baryon density.

In the chiral limit ($m_q=0$) considered here, where the dynamical mass also vanishes ($M=0$) in the QSP phase at high chemical potential, estimates~\cite{Matsuoka:2018huq} that rely on a non-zero current/constituent mass $m_q$ in the Bohr magneton factor $e/(2m_q)$, are inapplicable. For massless ultrarelativistic quarks, the relevant energy scale setting the electromagnetic response near the Fermi surface is the flavor chemical potential $\mu_f = (\pi^2 \rho_f)^{1/3}$. By replacing the non-relativistic magneton with the relativistic Fermi-surface magneton $\mu_{\text{eff}}^{(f)} = q_f / (2\mu_f)$ and incorporating the charge neutrality conditions $\rho_u = \rho_B + \rho_e$ and $\rho_d = 2\rho_B - \rho_e$, the spontaneous magnetization density in the chiral limit is expressed directly in terms of the baryon and electron densities:
\begin{equation}
M_{\text{mag}} = \frac{e F}{6 \pi^{2/3} G_T \rho_B^{1/3}} \left[ \frac{2}{\left(1 + \frac{\rho_e}{\rho_B}\right)^{1/3}} + \frac{1}{\left(2 - \frac{\rho_e}{\rho_B}\right)^{1/3}} \right].
\label{eq:Mmag_chiral_neutral}
\end{equation}

\begin{figure}[h!]
\begin{center}
\includegraphics[scale=0.5]{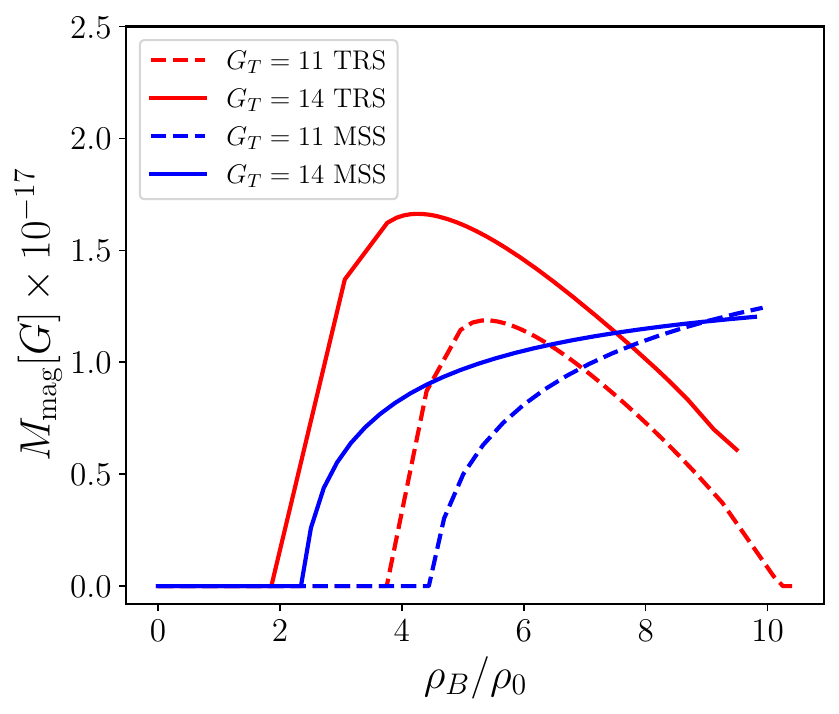}
\end{center}
\vspace{-0.5cm}
\caption{Magnetic moment of the hybrid star.}
\label{magnetic_moment}
\end{figure}

In Fig.~\ref{magnetic_moment}, we plot the magnetic moment strength versus baryon density, normalized by the normal nuclear density in both methods. We observe that the magnetic moment generated by the spin polarization phase in the MSS method is qualitatively and quantitatively different from that of the TRS. While the magnetic moment in the TRS method exhibits non-monotonic behavior and tends to decrease at $\rho_B = 8.5 \rho_0$ for $G_T = 11\text{ GeV}^{-2}$ and at $\rho_B = 9 \rho_0$ for $G_T = 14\text{ GeV}^{-2}$, the magnetic moment in the MSS method grows monotonically with the baryon density, enhancing the magnetic field generated by this phase. It is also observed that for stronger couplings, the magnetic moment grows faster in the MSS method. We obtain the magnetic moment of the order of $M_{\text{mag}} \sim 10^{17}~\text{ G}$, which is consistent with those observed in neutron star measurements.

\section{Conclusions and Perspectives} 
\label{sec:conclusion}

In this work, we investigated the effect of regularization on the appearance of a spin-polarized phase in high density quark matter using the two-flavor NJL model with a tensor interaction under $\beta$ equilibrium. We compared the traditional regularization scheme, in which the cutoff is applied directly to integrals containing both vacuum and medium-dependent quantities, with the medium separation scheme, in which algebraic subtractions isolate the divergent vacuum contributions and leave the finite medium-dependent terms cutoff independent.

Our results show that this distinction has significant physical consequences. In TRS, the QSP phase appears abruptly and disappears again at high chemical potential, leading to nonmonotonic thermodynamic behavior and, for some tensor couplings, acausal values of the speed of sound. In MSS, the QSP phase sets in at slightly higher chemical potential but grows monotonically with density. As a result, the thermodynamic observables are smoother, and the corresponding equation of state is softer than in TRS. The hybrid-star sequences obtained with MSS display different stability trends, and for the tensor couplings considered here, the masses are clearly affected by the QSP.  We also found that the magnetization associated with the QSP phase is strongly regularization dependent. While the TRS result decreases at high density, the MSS result grows monotonically with baryon density. These findings indicate that predictions for the magnetic and stellar properties of spin-polarized quark matter within a NJL-type model are, not surprisingly, highly sensitive to the treatment of medium contributions.

Extensions of the present analysis to include vector-channel interactions, color-superconducting pairing, and a potential interplay between chiral symmetry breaking and spin polarization constitute natural and well-motivated next steps. Vector interactions are known to increase the stiffness of the quark-matter equation of state, thereby facilitating the support of more massive hybrid-star configurations \cite{Fukushima:2008is,Sakai:2008ga,Zhang:2009mk,Bonanno:2011ch,Alford:2013aca,Denke:2013gha}. Furthermore, the present calculation was restricted to two-flavor quark matter in order to isolate the effect of the regularization procedure on the QSP phase. This restriction is nontrivial: at densities relevant for massive compact stars, the appearance of strange quarks and additional pairing patterns may become energetically favored, with concomitant modifications to the equation of state, the onset of spin polarization, and the predicted stellar configurations \cite{Rajagopal:2000wf,Rajagopal:2000ff,Fukushima:2008wg,Bonanno:2011ch,Morimoto:2020fkg,Kohyama:2016fif}. An additional direction is to assess the consequences of applying the MSS framework to the problem of rotation-induced magnetization \cite{Kiefer:2025xdp}. Investigations along these lines are currently underway.

\section*{Acknowledgements}
This work was partially supported by Conselho Nacional de Desenvolvimento Cient\'{\i}fico e Tecnol\'{o}gico (CNPq) grants nos. 313254/2025-7 (GK), 317048/2023-6 (MMF), 420896/2025-2 (MMF) and 312155/2023-9 (RC), Fundação de Amparo {\`a} Pesquisa e ao Desenvolvimento Cient\'{\i}fico e Tecnol{\'o}gico do Maranh{\~a}o (FAPEMA) grant nos. 00812/19 (RC) and 12299/22 (RC), and Funda\c{c}\~{a}o de Amparo \`{a} Pesquisa do Estado de S\~{a}o Paulo (FAPESP), grant nos. 2018/25225-9 (GK) and 2021/14335-0 (GK).

\bibliographystyle{apsrev4-1}
\bibliography{qsp_reg}


\end{document}